\documentstyle[12pt]{article}

\begin{document}

\title{The Initial Value Problem in Light of Ashtekar's Variables}{}{}

\author{
{\em Riccardo Capovilla} \thanks{Centro de Investigacion y de Estudios
Avanzados, I.P.N., Apdo. Postal 14-740, Mexico 14, D.F., Mexico.
{\it e-mail}: rcapovi@cinvesmx.
This author was partially supported by CONACyT Grant No. 1155-E9209.}
\and
{\em John Dell} \thanks{Thomas Jefferson High School for Science and
Technology, 6560 Braddock Road, Alexandria, VA 22312, USA.
{\it e-mail}: jdell@umdhep}
\and
{\em Ted Jacobson} \thanks{Department of Physics, University
of Maryland, College Park, MD 20742, USA. {\it e-mail}: jacobson@umdhep.
This author's research was
supported in part by the National Science Foundation under Grants No.
PHY91-12240 at the University of Maryland and PHY89-04035 at the Institute
for Theoretical Physics, Santa Barbara, California.}}

\date{} 

\maketitle
\vspace{-10pt} 

\begin{abstract}
The form of the initial value constraints in Ashtekar's hamiltonian
formulation of general relativity is recalled, and the problem of
solving them is compared with that in the traditional metric
variables. It is shown how the general solution of the four
diffeomorphism constraints can be obtained algebraically provided
the curvature is non-degenerate, and the form of the remaining (Gauss
law) constraints is discussed. The method is extended to cover
the case when matter is included, using an approach due to
Thiemann. The application of the method to vacuum Bianchi models
is given. The paper concludes with a brief discussion of
alternative approaches to the initial value problem in the
Ashtekar formulation.
\end{abstract}

\def\eg{{\it e.g.${~}$}}
\def\ie{{\it i.e.${~}$}}
\def\frac#1#2{{\textstyle{#1\over\vphantom2\smash{\raise.20ex
        \hbox{$\scriptstyle{#2}$}}}}}

\section{Introduction} 

It is with great pleasure that we dedicate this paper to Dieter Brill,
our teacher, advisor, and colleague, on the occasion of his 60th birthday.
Our contribution concerns the initial value problem for general relativity,
which is amongst Dieter's many areas of expertise. As is the case with
almost all research activity developed around the general relativity group
at the University of Maryland, the ideas we will present have benefitted from
Dieter's always kind and sometimes maddening insightful questioning. Of
course
it is our wish that this paper will prompt some more such questioning.

General relativity is invariant under four dimensional diffeomorphisms.
In the hamiltonian formulation, this invariance manifests itself
in the presence of constraints on the canonical variables.
If the constraints are satisfied at one instant of time,
they continue to hold at all times. The initial
value problem is the problem of giving a construction for
a parametrization of the general solution of the constraints.

The most well-developed approach to the initial value problem is based on
the ``conformal technique" (see for example \cite{CBY} and
references therein). In this approach, the phase space variables for
general relativity are the 3-metric $q_{ab}$ and its conjugate momentum
$p^{ab}$ which is a tensor density of weight 1, closely related to the
extrinsic curvature of the spatial hypersurface in a solution to the
field equations. There are four constraints, consisting of
one Hamiltonian constraint, which generates diffeomorphisms
normal to the spacelike surface on which the initial data are
given, and three momentum constraints, which generate
spatial diffeomorphisms. The Hamiltonian constraint is viewed
as a (quasi-linear,
elliptic) equation determining the conformal factor of the 3-metric. The
momentum constraints determine the ``longitudinal" piece of
$p^{ab}$.
In this approach, the freely specified data are, in
principle, the conformal equivalence class of the 3-metric,
and the transverse traceless part of the momentum $p^{ab}$.

The conformal technique is well suited to addressing
questions of existence and uniqueness of solutions to the
field equations. However, for two reasons it is not suited
to reduction of the phase space to the unconstrained degrees of
freedom. First, the ``freely specifiable" transverse
traceless tensor densities are not known for an arbitrary
conformal 3-metric. Second, the solutions to the constraint
equations determining the conformal factor and
longitudinal part of the momentum are not given explicitly.

An alternative representation for hamiltonian general relativity
was introduced in the mid-eighties by Ashtekar
\cite{Ashtekar1}. The Ashtekar representation may be
understood (and was discovered originally) as resulting from
a complex canonical transformation that goes from the
(triad-extended) ADM gravitational phase space variables to a new set
of variables. The new variables are a spatial SO(3,C)
connection, and an {\it so}(3,C)-valued vector density  as its conjugate
momentum.
The major benefit of this representation is that
the constraints take a simpler form in terms of these variables
than in the ADM representation. First, the constraints turn
out naturally to be polynomial of low order in the Ashtekar
phase space variables.
Moreover, for the vacuum case, the scalar constraint
is homogenous in the Ashtekar
canonical momenta. This simplification has allowed
considerable progress in many questions of gravitational
physics. Perhaps the most impressive results have been
obtained in the quantum theory,
where this formulation has made it possible to perform the
first few steps in the Dirac quantization of general
relativity \cite{JacSmo,RovSmo,Ashtekar2}.

Given that the Ashtekar representation simplifies the form of the
constraints, it is natural to reconsider the classical initial
value problem in this context. As it turns out, generically
in phase space, it is possible to obtain {\it algebraically}
the general solution to the Ashtekar version of the diffeomorphism
constraints of general relativity!
\cite{CDJ1,TT}
However, this does not give an easy solution of the initial
value problem.
The reason is that, due to the covariance of the Ashtekar formalism
under local SO(3,C) rotations, there are three additional constraints
whose form is identical to the non-abelian form of Gauss' law familiar
from Yang-Mills theories. Once the diffeomorphism contraints are solved by
the method to be described below, one still has to face the Gauss law,
which has now become a non-linear first order partial differential
equation on the remaining variables.
In addition, the reality conditions restricting to the phase space of
real, Lorentzian GR need to be enforced.

Still, one can hope that the relocation of the difficulties
afforded by the Ashtekar approach may prove to be fruitful
in some applications. This already seems to be the case in
the context of GR reduced by symmetry conditions. In addition,
the interplay between ``gauge-fixing" and solving the constraints
is different in the Ashtekar approach, which is a fact that remains to
be fully explored.

The rest of this paper is organized as follows.
The form of the initial value constraints in Ashtekar's hamiltonian
formulation of general relativity is recalled in section
2, and in section 3 it is shown how the general
solution of the four diffeomorphism constraints in vacuum (or with
a cosmological constant) can be obtained
algebraically provided the curvature is non-degenerate.
The form of the remaining (Gauss law) constraints is also
discussed, as is the relation between the solution given and the
structure of the 4-dimensional curvature. In section 4
the method is extended to cover
the case when matter is included, using an approach due to
Thiemann. The application of the method to vacuum Bianchi models
is given in section 5, and the paper concludes in section
6 with a brief discussion of alternative approaches to the
initial value problem in the Ashtekar formulation.

\section{Ashtekar's variables}
\label{2}

Ashtekar's representation of hamiltonian general
relativity has been the subject of extensive reviews
(cf. \cite{Ashtekar2,Rovelli}), so we shall just recall its main features.

The Ashtekar canonical coordinates are an SO(3,C) spatial connection,
$A_a^i$, and an {\it so}(3,C)-valued vector density of weight 1,
$E^a_i$. The fundamental Poisson bracket is given
by $\{A^i_a(x),E^b_j(y)\}=i\delta_j^i\delta_a^b\delta^3(x,y)$.
(Our notation
is as follows. Latin letters from the beginning of the alphabet
denote spatial indices, \eg $a, b, ... = 1, 2 , 3$. Latin letters from the
middle of the alphabet are SO(3,C) indices, $i, j,... = 1,2,3$. They are
raised and lowered with the Kronecker delta $\delta^{ij}$ and $\delta_{ij}$.
We will also use the totally antisymmetric symbol $\epsilon_{ijk}$,
with  $\epsilon_{123} = 1$. We use units with $c=G=1$.)

The vector density $E^a_i$ may be
identified with the densitized spatial triad $\sqrt{q} e^a_i $, with the
contravariant spatial metric given by $q^{ab} = e^a_i e^{bi} $. In turn, the
SO(3,C) connection may be identified in a solution with the spatial pullback
of the self-dual part of the spin-connection.

These variables parametrize the phase space of {\it complex} general
relativity. A real metric with Lorentzian signature may be recovered
by imposing appropriate reality conditions (cf. \cite{Ashtekar2}).
These conditions amount to the requirement that $E^a_i E^{bi} $ be real,
and that its time derivative be real. If these conditions are satisfied
initially then the dynamical evolution will preserve them in time.
A metric of Euclidean signature is obtained by simply taking
$A_a^i$ and $E^a_i$ real.

In terms of these variables, the constraints of (complex)
general relativity take the form
\begin{eqnarray}
\varepsilon_{abc} \epsilon^{ijk} E^a_i E^b_j B^c_k &=& -\rho
\label{eq:scalar}\\
\varepsilon_{abc} E^b_i B^{ci}  &=& -iJ_a \label{eq:vector} \\
D_a E^a_i &=& K_i\; . \label{eq:gauss}
\end{eqnarray}
Here $\varepsilon_{abc}$ is the standard Levi-Civita tensor
density. $B^a_i$ is the ``magnetic field" of the connection $A_a^i$,
defined by $B^a_i := \varepsilon^{abc} F_{bci} $, and
$F_{ab}^i := \partial_a A_b^i -  \partial_b A_a^i +
\epsilon^i{}_{jk} A_a^j A_b^k $. $D_a$ is the SO(3)-covariant derivative
determined by $A_a^i$. $\rho/\sqrt{q}$ and $J_a$ are (up to coefficients)
the matter energy and momentum densities respectively.
$K_i$ is a spin density, present only for half-integer spin matter fields.

In the following, we will call the constraints
(\ref{eq:scalar}) and (\ref{eq:vector}) ``diffeomorphism
constraints". The first generates diffeomorphisms normal to
the spatial hypersurface $\Sigma$
together with some SO(3) rotation, so it takes the place in Ashtekar's
formalism of the ADM hamiltonian constraint.
It will be called here the ``scalar constraint".
The second generates diffeomorphisms tangential to $\Sigma$
together with some SO(3) rotation, so it takes the place of the ADM momentum
constraint. It will be called here the ``vector constraint".

There are 3 additional constraints, (\ref{eq:gauss}),
of the Gauss-law type familiar from Yang-Mills theories.
They generate local SO(3,C) rotations, under which the Ashtekar formalism
is covariant.

The constraints are polynomial in the gravitational
phase space variables and for scalar
and spin-1/2 matter. For Yang-Mills type fields, polynomiality is mantained
only if one multiplies the scalar constraint (\ref{eq:scalar}) through
by $det q=det E^a_i$, which is cubic in $E^a_i$.
Moreover, it is remarkable that the gravitational contribution to the
scalar constraint (\ref{eq:scalar})
is {\it homogenous} in the momenta $E^a_i$. This should
be compared with the ADM Hamiltonian constraint,
where the term quadratic in the
canonical momenta must be balanced by a ``potential" term given by the
3D Ricci scalar times the determinant of the 3-metric.
Note that while the vector and Gauss constraints (\ref{eq:vector})
and (\ref{eq:gauss}) are densities of weight 1, the scalar constraint
(\ref{eq:scalar}) is of weight 2.

\section{Algebraic solution of the vacuum diffeomorphism constraints}
\label{3}

We now proceed to show how one can use the simplification
of the constraints for general relativity provided by the
Ashtekar formalism to tackle the initial value problem.
In particular, we will show how the general solution of the
the scalar and vector diffeomorphism constraints can be obtained
by algebraic methods. We originally discovered this solution
in the context of a Lagrangian pure spin-connection formulation
of GR, in which one solves for the metric variables (self-dual
2-forms) in terms of the connection \cite{CDJ1}. However, it is
unnecessary to view the technique in that context.

The first step is to assume that the magnetic field $B^a_i $
is non-degenerate as a $3\times3$ matrix.
In a generic real, Lorentzian spacetime, the real and imaginary parts of
the complex equation $det B=0$ will define two surfaces, and their
intersection
will give a one-dimensional submanifold on which $B^a_i$ is degenerate.
That is, generically $B^a_i$ is non-degenerate except on a set of measure
zero. (The points in phase space with $det B = 0$ {\it everywhere}
form a set of measure zero in phase space. The identity of these
points is best seen in the covariant formalism (cf. \cite{CDJ1,CDJ3}).
It turns
out that they correspond to space-time metrics
of Petrov type 0, (4), (3,1), when $E^a_i$ is non-degenerate.)
We will ignore here any problems that might arise as a consequence of
degeneracy of $B^a_i$.

Now, since $B^a_i$ is assumed to be non degenerate, one
may use it as a basis for the space of SO(3,C)
vector densities. In particular,
one can expand the momentum $E^a_i$ with respect to it,
\begin{equation}
E^a_i = P_ {ij} B^{aj}\; ,
\label{eq:expansion}
\end{equation}
for some (non-degenerate) $3\times3$ matrix $P_{ij}$.

We first consider the vacuum case, $\rho=0$, $J_a=0$, $K_i=0$.
It is easy to see that the vector constraint (\ref{eq:vector})
implies that $P_{ij}$ must be symmetric. The scalar constraint
(\ref{eq:scalar}) implies that $P_{ij}$ must satisfy a
quadratic algebraic condition,
\begin{equation}
(P^i_i )^2  - P_{ij} P^{ij} = 0\; .
\label{eq:condition}
\end{equation}
This condition fixes one of the six components
of $P_{ij}$ with respect to the others.

We know of two methods for solving the scalar constraint (\ref{eq:condition}).
In the first method, one solves (\ref{eq:condition}) for the trace of $P$.
Decomposing $P_{ij}$ into the trace $TrP$ and tracefree part
$\hat{P}_{ij}:=P_{ij}-\frac{1}{3}TrP \delta_{ij}$, (\ref{eq:condition})
becomes
the condition
\begin{equation}
Tr P = \pm \bigl(\frac{3}{2} Tr \hat{P}^2\bigr)^{1/2}\; .
\label{eq:TrP}
\end{equation}

In the second method for solving (\ref{eq:condition}), one notes that the
characteristic equation for $P$ implies that
$Tr P^{-1}=[Tr P^2-(Tr P)^2]/2detP$. Thus (\ref{eq:condition}) is equivalent
to
the tracelessness of $P^{-1}$, provided $P$ is invertible (which it must
be if both $B^a_i$ and the 3-metric are nondegenerate).
Let $\psi$ denote $P^{-1}$. Then the characteristic equation for
$\psi$, assuming $Tr \psi=0$, reads
$\psi^3-\frac{1}{2}(Tr \psi^2)\psi-det \psi I =0$, giving
$\psi^{-1}=[\psi^2-\frac{1}{2}(Tr \psi^2)I]/det \psi.$
Thus the general solution to (\ref{eq:condition}) for invertible $P$ can be
explicitly written in terms of the 5 independent
components of a tracefree $3\times3$ matrix $\phi$ in the form
\begin{equation}
P=\phi^2-\frac{1}{2}(Tr \phi^2)I\; .
\label{eq:phi}
\end{equation}
One is left with the Gauss constraint (\ref{eq:gauss}), which becomes
\begin{equation}
B^a_i D_a P^{ij} = 0\; ,
\label{eq:gauss2}
\end{equation}
where we have used the Bianchi identity $D_a B^a_i = 0$.
In view of (\ref{eq:TrP}) or (\ref{eq:phi}), the Gauss constraint
becomes a {\it non-linear} equation in $\hat{P}_{ij}$ or $\phi$
respectively. It is here, and in the reality conditions, that the
initial value problem now resides.

We can consider the Gauss law constraint as 3 conditions on
the 5 independent components of $P^{ij}$, for a
fixed $A_a^i$. Then there are presumably 2 free functions worth
of solutions for this equation, yielding the two indepenent metric degrees
of freedom of GR. Up to this stage the connection $A_a^i$ has remained
entirely
arbitary. By use of the 4 parameter diffeomorphism and 3 parameter
SO(3) transformations, one could now fix all but 2 of the 9 components
of the connection, yielding the other half of the coordinates on the
reduced phase space.

If a spherically symmetric ansatz is assumed, then the above method of
solving the constraints can be carried out entirely,
and the Gauss constraint reduces to an ordinary differential equation
that can be solved \cite{Bengspher}.

The existence in general of solutions to the Gauss
law constraint (\ref{eq:gauss2}) with $P^{ij}$ restricted by (\ref{eq:TrP}) or
(\ref{eq:phi}) is a problem that has not been addressed. To indicate
just one complication that can arise, consider the asymptotically flat case.
If $A_a^i$ is asymptotically
the spin-connection corresponding to a negative-mass spacetime,
then we know by the positive energy theorem that it must be impossible to
find a regular solution to (\ref{eq:gauss2}) for $P^{ij}$,
subject to the reality conditions.

The geometrical interpretation of the matrix $P^{ij}$
is easily seen from the covariant point of view \cite{SD2F}. The densitized
triad $E^{ai}$ of the Ashtekar formalism is related to the triad of
anti-self-dual 2-forms by $E^{ai}=\epsilon^{abc}\, \Sigma_{bc}^i$ \cite{MF}.
Now the curvature 2-form $R^i$ of the spin-connection can be expanded
in terms of the self-dual and anti-self-dual 2-forms as
as $R_i=\psi_{ij}\, \Sigma^j + \frac{1}{3}\Lambda\, \Sigma_i +
\Phi_{i\bar{j}}\bar{\Sigma}^{\bar{j}}$.
$\psi_{ij}$ is symmetric and tracefree and is just
the Weyl spinor in SO(3) notation. $\Lambda$ is proportional to
the Ricci scalar, and $\Phi_{i\bar{j}}$ is equivalent to
the tracefree part of the
Ricci tensor. Thus in vacuum, $\Lambda$ and $\Phi_{i\bar{j}}$ vanish,
and one can solve for the anti-self-dual 2-forms in terms of the curvature as
$\Sigma^i=(\psi^{-1})^{ij}\; R_j$. The dual of the spatial pullback of this
equation yields immediately the general solution given above for
the four diffeomorphism constraints in Ashtekar's formalism,
with $P^{ij}$ identified with $(\psi^{-1})^{ij}$.

Let us now consider the addition of a cosmological
constant $\Lambda$. This modifies only the
scalar constraint, which becomes
\begin{equation}
\varepsilon_{abc} \epsilon^{ijk} E^a_i E^b_j B^c_k -
(1/3) \Lambda \varepsilon_{abc} \epsilon^{ijk} E^a_i E^b_j E^c_k = 0\; .
\label{eq:cc}
\end{equation}
One may now proceed as in the vacuum case. The only difference is
that the algebraic condition (\ref{eq:condition}) is replaced by
\begin{equation}
(P^i_i )^2  - P_{ij} P^{ij} =  2 \Lambda (det P)\; .
\label{eq:cond2}
\end{equation}
This can be regarded as a cubic equation for $TrP$.
It is equivalent to the statement that the inverse
of $P$ has trace equal to $\Lambda$.
In this case, P corresponds to the inverse of the
matrix $(\psi_{ij}+\frac{1}{3}\Lambda\, \delta_{ij})$.

An interesting special solution of the constraints in the case
of a nonvanishing $\Lambda$ is given by the Ansatz $P_{ij} =
(3/\Lambda) \delta_{ij}$, or
\begin{equation}
E^a_i = (3/\Lambda) B^a_i\; ,
\label{eq:ansatz}
\end{equation}
which solves automatically all of the constraints. (The Gauss
constraint is satisfied as a consequence of the
Bianchi identity $D_a B^a_i = 0 $.) This Ansatz was first introduced
by Ashtekar and Renteln \cite{AshtekarRenteln}, who observed
that it gives rise to self-dual solutions of the Einstein
equation with a non-vanishing $\Lambda $.

\section{Matter couplings}
\label{4}

It is possible to
extend the method just described to solve the diffeomorphism
constraints in the presence of matter using an
approach that has recently been brought to our attention by Thomas
Thiemann \cite{TT}. Thiemann's method proceeds as follows.

One begins with the constraints in the form
(\ref{eq:scalar}), (\ref{eq:vector}), (\ref{eq:gauss})\cite{ART}.
Then expanding $E^a_i=P_{ij} B^{aj}$ as in
(\ref{eq:expansion}), one finds that the vector constraint (\ref{eq:vector})
implies that the
anti-symmetric part of $P_{ij}$ no longer vanishes, but is given
instead by
\begin{equation}
A_{ij}:=P_{[ij]}={i\over 2B}\; \epsilon_{ijk} B^a_k\, J_a\; ,
\label{eq:A}
\end{equation}
where $B:=det B^a_i$. This is an explicit expression for $P_{[ij]}$
provided $J_a$ does not depend upon $P_{[ij]}$ as well. Since $J_a$
is just the generator of spatial diffeomorphisms for the matter
variables, it does not depend on the gravitational field variables
for the case of integer spin matter. In the spin-1/2 case, the diffeomorphism
is accompanied by an SO(3) rotation, which involves the spin connection
but not $E^a_i$. In all cases therefore,
$J_a$ is independent of $P_{ij}$. (In particular, in
the scalar, spin-1/2, and Yang-Mills cases, the structure of $J_a$
is $\pi\partial_a\phi$, $\pi^A D_a\psi_A$, and $e^{bI}f_{abI}$
respectively, where $e^{bI}$ and $f_{abI}$ are the
Yang-Mills-electric and (dual of) magnetic fields.)

Now turning to the scalar constraint, we begin by noting that
when $P_{ij}$ is decomposed into its symmetric and antisymmetric parts
$P_{ij}=S_{ij}+A_{ij}$, the left hand side of the scalar constraint
({\ref{eq:scalar}) is given
by
\begin{equation}
\epsilon_{ijk}\epsilon_{abc}E^{ai}E^{bj}B^{ck}
=det B\; [(TrS)^2-Tr S^2-Tr A^2]\; .
\label{eq:lhs}
\end{equation}

Let us first consider the case of a
single scalar field. Then we have
\begin{equation}
\rho=\pi^2+E^{ai}E^b_i\partial_a\phi\partial_b\phi+\epsilon_{ijk}
\epsilon_{abc}E^{ai}E^{bj}E^{ck}\; V(\phi)\; .
\label{eq:scalarrho}
\end{equation}
Now as in the vacuum case there are two approaches to solving the
scalar constraint. In the first method, one regards the
constraint as a cubic equation on the trace of $S_{ij}$, which can be
solved (albeit messily) in closed form. In the second approach, due
to Thiemann \cite{TT}, one notes that every term of the constraint is
either independent of the scalar field momentum $\pi$ or depends
on $\pi$ quadratically! To see why, observe that $J_a$ is linear
in $\pi$, and therefore according to (\ref{eq:A}), so too is $A_{ij}$.
Thus the
gravitational part of the constraint (\ref{eq:lhs}) is quadratic in
$\pi$. As for $\rho$ (\ref{eq:scalarrho}), the second (gradient) term is
independent of $\pi$ because, as one easily sees using (\ref{eq:A}),
$A_{ij}$ drops out of the expression
$E^{ai}\partial_a\phi=S^{ij} B^a_j \partial_a \phi$. Moreover, in the
third term one can show that $det E$ involves $A_{ij}$, and therefore
$\pi$, also only quadratically.

Thus one can simply solve the constraint for $\pi$ in closed
form. Even for the case $V(\phi)=0$, the resulting expression is
fairly complicated:
\begin{equation}
\pi=\pm\Biggl[\Bigl((TrS)^2-TrS^2+(S\xi)^2\Bigr)/
\Bigl(\frac{\xi^2}{2B}-1\Bigr)\Biggr]^{1/2}\; ,
\label{eq:pi}
\end{equation}
where $\xi^i:=B^{ai}\partial_a\phi$.

For the real theory, there is also the constraint that the
argument of the square root be a positive real number.
When $V(\phi)\ne 0$, the solution for $\pi$ becomes significantly
more complicated. It is unlikely that this is of any practical
use in general. However, Thiemann's intended application is
quantization of the spherically symmetric scalar-gravitational
system, for which the resulting expressions may be more useful.

Note that even when the argument of the square root
in (\ref{eq:pi}) is real, the solution for $\pi$ has a sign ambiguity.
This can lead to problems if one wishes to pass to the reduced phase
space for the purpose of quantization, since it is difficult to eliminate
the momentum and not lose part of the reduced phase space\cite{Tate}.

For other types of matter coupling the situation becomes yet
more complicated. For a massless spin-1/2 field, the scalar
constraint
remains quadratic (but not homogeneous)
in the two components $\pi^A$ of the spinor
field momentum, so one can in principle solve for one of these
components. In addition, the Gauss constraint is augmented by
a spinor field term in this case.
When Yang-Mills fields are included, the scalar constraint
remains polynomial only if it is multiplied through by $det E$,
resulting in a 4th order polynomial in the scalar field momentum.

\section{Bianchi models}
\label{5}

In this section, we apply to spatially homogenous vacuum space-times,
\ie Bianchi models, the method for the solution of the constraints
illustrated in section 3 for the
vacuum case. The hope is to gain some insight on how to deal
with the remaining, Gauss constraint in the form (\ref{eq:gauss2}).
In any case, this analysis may be of use in a minisuperspace
approximation of vacuum general relativity.

For Bianchi models, as follows from the assumption
of spatial homogeneity, this last constraint reduces
to an algebraic condition. There are some simplifications,
but the condition is still rather complicated. We
record it here for the benefit of inventory. One
interesting feature is that only the traceless
part of $P_{ij}$ enters in it.

The formulation of Bianchi models in terms of Ashtekar
variables was first worked out by Kodama \cite{Kodama}.
Here we follow the approach of Ashtekar and Pullin
\cite{AshtekarPullin}.

As is familiar from the ADM treatment of Bianchi models
(see \eg \cite{MacCallum,RyanShepley}), we consider a
kinematical triad of vectors, $X^a_I$, which commute
with the three Killing vectors on the
spatial hypersurface $\Sigma$. (Capital latin letters
$I,J,K,...$ will be used to label the triad vectors.)
The triad satisfies
\begin{equation}
[ X_I , X_J ]^a  = C_{IJ}{}^K  X^a_K\; ,
\label{eq:comm}
\end{equation}
where $C_{IJ}{}^K $ denote the structure constants
of the Bianchi type under consideration. The basis dual
to $X^a_I$, $ \chi_a^I$, satisfies,
\begin{equation}
2 \partial_{[a} \chi_{b]}^I = - C_{JK}{}^I \chi_a^J \chi_b^K\; .
\label{eq:comm2}
\end{equation}
Without loss of generality, one may set
\begin{equation}
C_{IJ}{}^K = \epsilon_{IJL} S^{LK} + 2 \delta_{[I}^K V_{J]}
\label{eq:decomp}
\end{equation}
with $S^{IJ}$ symmetric.
(This $S^{IJ}$ has nothing to do with the symmetric part $S^{ij}$
of $P^{ij}$ referred to earlier.)
{}From the Jacobi identities,
it follows that $S^{IJ} V_J = 0 $. The Bianchi classification
may be described in terms of the vanishing or not of $V_I$,
and the signature of $S^{IJ}$, subject to this condition.
The most popular models are Bianchi I, selected by $C_{IJ}{}^K = 0$,
and Bianchi IX, selected by $V_I = 0$  and $ S^{IJ} = \delta^{IJ}$.

The Ashtekar gravitational phase space variables may be expanded
with respect to the kinematical triad $X^a_I$, and $\chi_a^I$, as
\begin{eqnarray}
A_a^i &=& A_M^i \chi_a^M \\
E^a_i &=& det\chi\; E^M_i X^a_M
\end{eqnarray}
where $ det \chi $ denotes the determinant of $\chi_a^I$, which is
introduced in order to de-densitize $E^a_i$.
Similarly, the magnetic field may be expanded as
\begin{equation}
B^a_i = det\chi\; B^M_i X^a_M\; ,
\end{equation}
and $B^M_i$ is given by
\begin{equation}
B^M_i = - \epsilon^{MNP} C_{NP}{}^Q A_{Qi}
+ \epsilon^{MNP} \epsilon_{ijk} A_N^j A_P^k\; .
\label{eq:magnetic}
\end{equation}
The gravitational phase space has thus been reduced to the
matrices $\{ A_M^i , E^M_i \} $, \ie from
18 degrees of freedom per space point to only 18 in total.

The constraints for vacuum reduce to
\begin{eqnarray}
\epsilon_{MNP} \epsilon^{ijk} E^M_i E^N_j B^P_k &=& 0
\label{eq:scalarhom}\\
\epsilon_{MNP} E^N_i B^{Pi} &=& 0
\label{eq:vectorhom}\\
C_{KM}{}^K E^M_i + \epsilon_{ijk} A_M^j E^{Mk} &=& 0
\label{eq:gausshom}
\end{eqnarray}
where $B^M_i$ is given by (\ref{eq:magnetic}).

We can follow now the footsteps of section 3, for the solution
of these constraints.
Assuming that $B^M_i$ is non degenerate, we expand $E^M_i$ in terms of it:
\begin{equation}
E^M_i = P_{ij} B^{Mj}\; .
\end{equation}
{}From (\ref{eq:vectorhom}),
we find that $P_{ij}$ must be symmetric, and from (\ref{eq:scalarhom})
that
it must satisfy the algebraic condition (\ref{eq:condition}).
We arrive at the
last constraint, which now takes the form
\begin{equation}
C_{KM}{}^K B^{Mj} P_{ij} + \epsilon_{ijk} A_M^j B^M_l P^{lk} = 0\; .
\label{eq:gausshom2}
\end{equation}
Using (\ref{eq:magnetic}) and the Jacobi identities, after some algebraic
manipulations,
this may be written in the form
\begin{equation}
W^j \hat{P}_{ij} = Z_{ijk} \hat{P}^{jk}\; ,
\end{equation}
where $\hat{P}_{ij}$ corresponds to the traceless part of
$P_{ij}$, and
\begin{eqnarray}
W^j &:=& 3 V_Q \epsilon^{QMN} \epsilon^{jkl} A_{Mk} A_{Nl} \\
Z_{ijk} &:=& - 2 \epsilon_{ijl} A_M^l S^{MN} A_{Nk}\; .
\end{eqnarray}
Note that this condition involves only the traceless part of
$P^{ij}$.

We can now go through the Bianchi classification for
some understanding of this condition.
For a Bianchi model of type I, defined with $C_{IJ}{}^K = 0$,
this condition is trivially satisfied. If $S^{IJ} = 0$, which
defines a model of type V, the condition reduces to the
requirement that $W^i$ be a null eigenvector for $\hat{P}_{ij}$.
If $V_Q = 0$, one can look at the simple case
in which the connection matrix $A_{Mi}$ is
assumed to be diagonal (see Ref. \cite{AshtekarPullin}).
Then, recalling that without loss of generality $S^{IJ}$
can be put in diagonal form with $\pm 1$ or $0$ entries,
and letting $S^{IJ} = diag ( s_1 , s_2 , s_3 )$, the
condition takes the form
\begin{equation}
[s_1 (A_{11} )^2 - s_2 (A_{22} )^2 ] \hat{P}_{12} = 0\; ,
\end{equation}
together with its cyclic permutations. An obvious solution is
obtained if $\hat{P}_{ij}$ itself is diagonal.

\section{Other Approaches}
\label{6}

The introduction of Ashtekar's variables has generated
other approaches to the initial value problem in
general relativity. One is the recent
proposal by Newman and Rovelli \cite{NR}.
For a Yang-Mills theory they
use a Hamilton-Jacobi technique to solve Gauss' law.
The Yang-Mills physical degrees of freedom are coded
in one pair of scalar functions per dimension of the gauge
group, together with conjugate momenta. Each pair of scalar fields
defines a congruence of lines (``generalized
lines of force") by the intersections of their level surfaces.
The method can also be applied to gravity in
the Ashtekar formulation, where Newman and Rovelli solve not only the
Gauss constraint, but also the vector constraint, by the device
of using three of the scalar fields as spatial coordinates and letting
the remaining fields be given as functions of these.
It is not known at present if also the scalar constraint can be solved
using this technique.

Another potential avenue to the initial value problem is the
application of the Goldstone-Jackiw solution \cite{GoldstoneJackiw}
of the Gauss constraint in the SU(2) Yang-Mills theory to gravity in
the Ashtekar formulation \cite{Manojlovic}.
In this approach, one would start by solving the Gauss law.
The solution could then be used in the
diffeomorphism constraints. For Yang-Mills,
this procedure is not particularly useful,
because it results in a complicated
hamiltonian. For gravity, the jury is still out.

Finally, Thiemann \cite{TT} has just introduced a new approach
that enables him to solve {\it all} of the constraints.
The key idea is to start with the ansatz
$E^{ai}=\epsilon^{abc}D_{[b}v_{c]}^i$.
This ansatz involves no loss of generality provided the curvature
is non-degenerate, or provided the spin density
$K_i$ vanishes (so that the Gauss constraint implies $D_a E^{ai}=0$).
Using this ansatz, the Gauss constraint becomes a linear
condition on $v_a^i$.
The vector and scalar constraints are then solved in the presence of
matter by solving for some of the matter momenta.

\bibliographystyle{plain}

\end{document}